\documentclass[preprint,12pt]{elsarticle}

\usepackage{epsfig}

\usepackage{amssymb}

\journal{arxiv.org}

\begin{document}

\begin{frontmatter}
\title{Modeling the Propagation, Breaking and Drift of\\ Ocean Surface Wave}


\author{Jin-Liang Wang}

\address{Research Institute for ESMD method and Its Applications,\\
College of science, Qingdao University of Technology, Shandong, P.R. China, 266520.\\
 {E-mail: wangjinliang0811@126.com}}

\baselineskip = 23 pt

\begin{abstract}
 A new model other than the classical ones given by Airy, Stokes and Gerstner for the ocean surface wave is constructed. It leads to new understandings for the wave mechanisms: (1) A wave with bigger amplitude or smaller steepness travels faster; (2) The wave breaks when the front angle is bigger than 46.3 degree; (3) The magnitude of the wave drift should be smaller than that of the known Stokes drift.
 \end{abstract}


\end{frontmatter}



\baselineskip = 23 pt

\section{Introduction}
\setcounter{equation}{0}
The study of water wave is one of the oldest branches of hydrodynamics.
It can be dated back to the year 1687 when Newton
did an experiment with U-tube and
got the result ``the frequency of deep-water waves must be proportional
to the inverse of the square root of the wave length''.
As reviewed by Craik (2004), the classical wave theories were mainly developed
by the scientists from France, Germany and Britain
in the eighteenth and early nineteenth centuries.
Among all of them, the representative works are given
by Airy (1845) for linear wave, Stokes (1847) for nonlinear wave,
 Gerstner (1802) for cycloid wave and Earnshaw (1847) for solitary wave.
 After that time, the progresses are
 under the existing framework and on the wave-breaking investigation (Banner, 1993),
 the wind-wave growing mechanism (Phillips, 1957; Miles, 1957; Janssen, 2009),
 the wave-spectrum construction (Phillips, 1977; Wen and Yu, 1984)
 together with its applications in numerical ocean-wave forecast (Cavaleri et al, 2007; Mitsuyasu, 2002).
 One can also refer to the special issue ``Ocean Surface Waves''
 of \emph{Ocean Modeling}, Vol.70 (2013) for the latest developments on these aspects.
 The present article aims at constructing a new model other than those
 given by Airy, Stokes and Gerstner. As for the solitary wave on shallow water
 given by Earnshaw, it is beyond the topic of periodic wave in deep water and is omitted here.
 The default form of water wave is the so-called ``gravity wave'' on the ocean surface.

To carry out the new modeling, it needs some reviews on the classical wave models.

\subsection{On the Linear Wave Model}

 The classical linear wave theory in nowadays textbooks, such as those by
 Andersen and Frigaard (2011), Soloviev and Lukas(2006) and
Dean and Dalrymple (1991), mostly follows from the model of Airy (1845).

 As the problem concerned, the default model
should be the inviscid and incompressible Navier-Stokes equations.
 But the solving of these equations involves in determining
the upper surface boundary condition which is just the wave to look for (Stewart, 2005).
This nonlinear characteristic makes the problem insoluble in essence.
So, the classical results for surface waves are merely some kind of approximations and
the linear wave is the simplest one.
In fact, even the trigonometric wave forms themselves are some kind of conjectures.
Here the Cartesian coordinate system is adopted and only the 2-dimensional case is concerned.
The origin is chosen at the motionless water level with $x$ and $z$ pointing to the propagating direction
and the upward direction separately.

On the assumption that the amplitude $A$ is infinitely small
relative to the wave-length $\lambda$ ($=2\pi/k$), that is, $\varepsilon=Ak\ll 1$ and
the upper boundary can be almostly seen as a fixed flat surface,
 there is a linear approximation for the problem.
At this time, the surface traveling wave can be conjectured in the simplest trigonometric form:
\begin{equation}
\xi(x,t)=A\sin(kx-\omega t).
\end{equation}
 For the deep-water case with irrotational hypothesis on the flow,
 the solving of the simplified Navier-Stokes equations yields depth-dependent profiles
 for the wave and pressure:
 \begin{eqnarray}
 &&\xi^*(x,z,t)=Ae^{kz}\sin(kx-\omega t),\\
 &&P(x,z,t)=P_0+\rho g\left[Ae^{kz}\sin(kx-\omega t)-z \right]
 \end{eqnarray}
together with a relation:
\begin{equation}
\omega^2=gk,
\end{equation}
which is known as the ``dissipation relation''.
$P_0$ and $\rho$
are the constant air pressure on the surface and the water density.
Here we have omitted the horizontal and vertical velocities $u$ and $w$
which can be also outputted from the solving process.

By aid of dissipation relation, the wave-speed is derived to be
\begin{equation}
c=\frac{\lambda}{T}=\frac{\omega}{k}=\sqrt{\frac{g}{k}}=\frac{g}{\omega},
\end{equation}
here $T$ is the period which is related to frequency by $T=2\pi/\omega$.
This formula indicates that a wave with lower frequency or lower wave-number should travel faster.

Besides the limitation $Ak\ll 1$, there is another shortcoming for the linear wave model:
 \emph{It does not break and its theory is not suitable for wave-breaking problem}.

\subsection{On the Stokes Wave Model}
In case $\varepsilon=Ak$ ($<1$) is not infinitely small,
there is a finite-amplitude Stokes wave model owing to Stokes (1847). For this case,
 the surface wave and dispersion relation for the deep-water case is approximated by the asymptotic expansion technique:
 \begin{eqnarray}
 &&\xi(x,t)=A\cos{\theta}+\frac{1}{2}\varepsilon A\cos{2\theta}
 +\frac{3}{8}\varepsilon^2A\cos{3\theta}+\cdots,\\
 &&\omega^2=\left[1+\varepsilon^2+\frac{5}{4}\varepsilon^4+\cdots \right]gk
 \end{eqnarray}
 with $\theta=kx-\omega t$ (Soloviev and Lukas, 2006; Wen and Yu, 1984; Stewart, 2005).
  The corresponding pressure profile
 can be approximated by:
 \begin{eqnarray}
 P(x,z,t)=P_0+\rho g\left[e^{kz}\xi(x,t)-z \right].
  \end{eqnarray}

 Relative to the linear wave, the Stokes wave looses the request on the wave steepness $\varepsilon=Ak$
 and it yields three brand-new results: (1) It accords well with the actual one which has
 sharp crests and flat troughs; (3) The asymmetry of crest and trough lifts the equilibrium
 to the height $\varepsilon A/2$ higher than the motionless water level;
  (3) The particle's trajectory is not a closed circle,
 the water body has a mean drifting velocity (known as ``Stokes drift''):
 \begin{eqnarray}
 U_s=\varepsilon^2 c e^{2kz_0},
  \end{eqnarray}
here $z_0$ is the water depth with a negative sign. The surface Stokes drift accords with the case $z_0=0$.

 For a too big steepness $\varepsilon$ which accords with
 a too sharp crest the Stokes wave may break down
 (Massel, 2007). So it is suitable for wave-breaking problem and
 the criteria given by Stokes are seen as classical ones nowadays.
  The shortcoming lies in the irrotational-flow hypothesis.

\subsection{On the Gerstner Wave Model}

On the assumption that the particle's trajectory is a circle,
Gerstner (1802) found a rotational cycloid wave
(known as ``Gerstner wave'').
It is an exact solution of Lagrangian form in deep water (Soloviev and Lukas, 2006):
\begin{eqnarray}
\left\{
\begin{array}{ll}
x(x_0, z_0, t)=x_0-Ae^{kz_0}\cos(kx_0-\omega t),\\[2mm]
z(x_0, z_0, t)=z_0-Ae^{kz_0}\sin(kx_0-\omega t),
\end{array}
\right.
\end{eqnarray}
here $(x_0, z_0)$ denotes the equilibrium of the water particle
which can be seen as a location shift from that for initial time $t=0$
in the strict Lagrangian frame.
We note that, for convenience of comparison with the following new model,
a translation is done on the phase angle by $\pi/2$.

To ensure the pressure at the free surface $z=\xi(x,t)$ be a constant (the dynamic boundary condition),
it requires a dispersion relation
$\omega^2=gk$ same to the linear wave. For this case,
the water pressure should satisfy (Wen and Yu, 1984):
\begin{eqnarray}
P=P_0-\rho g z_0-\frac{1}{2}\rho g A^2 k\left(1-e^{2kz_0} \right)
\end{eqnarray}
which has noting to do with the variables $x_0$ and $t$. That is,
the water pressure is merely in the depth-dependent form $P(z_0)$
 and does not change during the process of wave motion.
It is very special. Our common sense is that:
To support the $t$-periodic wave motion, the pressure should also vary
in a $t$-periodic manner, such as in eqns.(1.3) and (1.8).

Relative to the Stokes wave, its advantages lie in the concise expression and
the abandon of irrotational hypothesis.
To some extent, it accords better with the actual one which has sharp crests and flat troughs.
However, its descriptive power is limited, after all, not all waves have circular particle trajectories.
To accord with the physics, it need reforming.

\section{Remodeling the Wave Motion}

\setcounter{equation}{0}
From the previous analysis we know the traditional approaches given by Airy, Stokes and Gerstner
take the conjectured wave forms as the preconditions or
approximate it with asymptotic expansion technique.
What is more, the water pressures are given as corollaries in the last.
Here we take an inverse approach to do it. Let the wave model be the object, the conjecture is done on the pressure.

To make modeling here we follow the fundamental assumptions adopted by the classical models:
 (1) the fluid is inviscid and incompressible;
 (2) the effect of surface tension is neglectable;
 (3) the particle at the water surface always maintains at it during
 the reciprocating process;
 (4) the water is deep relative to the concerned wave.

Take one water particle as the research object, we describe it
by Lagrangian coordinates $(x(x_0,z_0,t), z(x_0,z_0,t))$. Its motion satisfies the equations (Price, 2006):
 \begin{eqnarray}
a_x=-\frac{1}{\rho}\frac{\partial P}{\partial x},\qquad
a_z=-\frac{1}{\rho}\frac{\partial P}{\partial z}-g,
\end{eqnarray}
 here $a_x$ and $a_z$ are the simple denotations of the accelerations ${\partial^2 x}/{\partial t^2}$
 and ${\partial^2 z}/{\partial t^2}$. It requires
 the knowledge of water pressure.

\subsection{Conjecture on the Pressure}

For a hydrostatic case, in case the density can be seen as
a constant, the water pressure $P$ increases linearly along with
the water-layer thickness $s$, that is, $P=P_0+\rho g s$.
When the fluid is flowing with a wave motion designated by $z=\xi(x,t)$ on the upper surface,
the water pressure roughly obeys this rule with
\begin{equation}
P\approx P_0+\rho g [\xi(x,t)-z],
\end{equation}
this is the so-called ``quasi-hydrostatic approximation''
usually adopted in physical oceanography (Stewart, 2005). As the problem concerned,
if this kind of approximation is adopted,
then it follows from eqn.(2.1) that $a_z\approx 0$. This means the vertical velocity almost keep unchanged.
It is impossible! The common sense is that the vertical velocities at the crest and the trough are all zero
but those at the mean level are not zero.

Besides the quasi-hydrostatic approximation there is another case, might as well, call it by
``gravitational approximation''.
As analyzed in the previous section, if the gravity is the main restoring force for the wave motion,
then there should be $a_z\approx -g$ as the particle at the crest part. For this case, $\partial P/{\partial z}\approx 0$.
This means there is no relative vertical force between two arbitrary water layers.
Hence, the horizontal pressure gradient force due to the slant water body is empty which leads to $a_x\approx 0$.
This is also a strange case.

As analyzed above, the ``quasi-hydrostatic approximation'' and the ``gravitational approximation'' are two extreme cases:
the vertical pressure gradient force is too strong for the first case and too weak for the second case.
With this understanding, we conjecture the pressure in an eclectic way.
Notice that the water pressures (eqns.(1.3) and (1.8)) for the linear wave and Stokes wave are deduced from the
 Navier-Stokes equations, their forms should be more objective
and have more reference values. Enlightened by these we estimate the pressure by
 \begin{eqnarray}
P=P_0+\rho g\left[e^{kz_0}\xi(x,t)-z \right].
\end{eqnarray}
This expression is a modified version with $z_0$ substituting $z$ in the exponential function.
For this case, the free surface $z=\xi(x,t)$ with equilibrium at $z_0=0$ accords well with the dynamic boundary condition $P=P_0$.
We note that the incorporating of $z_0$ here is permitted.
In fact, under the Lagrangian frame, the functions $x, z$ and $P$ can be all expressed by the variables $x_0, z_0$ and $t$.
 Yet, under the Euler frame whose variables are $x, z$ and $t$, it is strange to incorporate $z_0$ into eqns.(1.3) and (1.8).

\subsection{Approximations on the Accelerations}

It is a reasonable hypothesis that, during the moving process,
a small cubic water body keeps its shape from $(x_0,z_0)$ to $(x,z)$ (Price, 2006).
For this case,
\begin{eqnarray}
\frac{\partial P}{\partial z_0}=\frac{\partial P}{\partial x}\frac{\partial x}{\partial z_0}
+\frac{\partial P}{\partial z}\frac{\partial z}{\partial z_0}
=\frac{\partial P}{\partial x}\cdot 0
+\frac{\partial P}{\partial z}\cdot 1=\frac{\partial P}{\partial z}.
\end{eqnarray}
It follows from eqns.(2.1), (2.3) and (2.4) that
\begin{eqnarray}
a_x=-ge^{kz_0}\frac{\partial\xi}{\partial x},\qquad
a_z=-gke^{kz_0}\xi.
\end{eqnarray}
\emph{These mean the horizontal motion is due to the
pressure-gradient force caused by the slant water body
and the vertical motion is due to the variation of the surface elevation
(can be understood as the variation in the previous period,
it squeezes the water body and leads to new vertical motion).}

To simply the horizontal acceleration, we take the wave slope $\partial\xi/{\partial x}$ as a parameter
and denote $\delta$ the average of its absolute value over a wave-length respect to
the fixed time $t=0$, that is,
\begin{eqnarray}
 \delta&=&\frac{1}{\lambda}\int_0^{\lambda}\left|\frac{\partial\xi(x,0)}{\partial x}\right|dx
\approx\frac{4}{\lambda}\left|\int_0^{\lambda/4}\frac{\partial\xi(x,0)}{\partial x}dx\right|\nonumber\\
&=&\frac{4}{\lambda}|\xi(\lambda/4,0)-\xi(0,0)|=\frac{4}{\lambda}|-A-0|=\frac{4A}{\lambda}=\frac{2Ak}{\pi},
 \end{eqnarray}
 here the position of wave trough is set on $x=\lambda/4$.
 We note that $\delta$ is the real wave steepness (wave slope),
 it relates to the common one $\varepsilon=Ak$ by $\delta=2\varepsilon/\pi$.

With this simplicity,
the horizontal acceleration can be further approximated by
\begin{eqnarray}
a_x\approx\left\{
\begin{array}{ll}
\delta g e^{kz_0},\qquad {\partial\xi}/{\partial x}<0,\\[2mm]
-\delta g e^{kz_0}, \;\quad {\partial\xi}/{\partial x}>0.
\end{array}
\right.
\end{eqnarray}

To simply the vertical acceleration, we approximate the time-variation of $\xi$ at $x=0$
by a piecewise linear function:
\begin{eqnarray}
\xi(0,t)\approx\left\{
\begin{array}{lll}
4At/T,\quad t\in [0,T/4]\\[2mm]
-A(4t/T-2),\quad t\in [T/4, 3T/4],\\[2mm]
A(4t/T-4),\quad t\in [3T/4, T].
\end{array}
\right.
\end{eqnarray}
With this approximation we get
\begin{eqnarray}
a_z\approx \frac{\pi}{2}\delta g e^{kz_0} \cdot\left\{
\begin{array}{lll}
-4t/T,\quad t\in [0,T/4]\\[2mm]
4t/T-2,\quad t\in [T/4, 3T/4],\\[2mm]
4-4t/T,\quad t\in [3T/4, T],
\end{array}
\right.
\end{eqnarray}
here the replacement is adopted on $Ak=\pi \delta/2$.

\subsection{Modeling the Motion of the Surface Particle}

In the following we consider the surface wave which accords with $z_0=0$.
Notice that it is a synthesis of transversal wave and longitudinal wave,
with the aid of the deduced accelerations, we decompose the motion into the vertical part and horizontal part and
model them separately.

To be simple, the start time $t=0$ is chosen from the undisturbed state with particle at $(0,0)$.
To rewrite $x(0,0,t)$ and $z(0,0,t)$ by $x(t)$ and $z(t)$,
then the upward motion can be understood as from a sudden impulse with initial velocity $(0,w_0)$.
We note that the effect of current is left out. In case $u(0)\neq 0$ there should be a
discussion on the wave-current interaction.

\subsubsection{Vertical Motion}

To describe the vertical motion distinctly we divide it into three stages:
(1) from the zero-level to the crest; (2) from the crest to the trough;
 (3) from the trough to the zero-level.

In view of eqn.(2.9) the upward velocity of the first stage satisfies
  \begin{eqnarray}
   w(t)=w_0-\int_0^t \frac{2\pi}{T}\delta g s ds=w_0-\frac{\pi}{T}\delta g t^2.
  \end{eqnarray}
Its corresponding displacement is
  \begin{eqnarray}
   z(t)=\int_0^t w(s)ds=\int_0^t \left[w_0-\frac{\pi }{T}\delta g s^2\right]ds
   =w_0 t-\frac{\pi}{3T}\delta g t^3.
  \end{eqnarray}
  Particularly, when the surface particle attains the crest with $w(T/4)=0$ and $z(T/4)=A$, there should be
  $w_0=\pi\delta g T/16$ and
  \begin{eqnarray}
   A=\frac{\pi}{16}\delta g T\cdot \frac{T}{4}-\frac{\pi}{3T}\delta g \left(\frac{T}{4} \right)^3=\frac{\pi}{96} \delta g T^2.
  \end{eqnarray}
  which yields the relation for $T$, $A$ and $\delta$:
  \begin{eqnarray}
  T=4\sqrt{\frac{6A}{\pi g\delta}}.
  \end{eqnarray}
  This formula can be used to guide the observation at sea.
  For example, if the observed amplitude and slope are $A=2$m
  and $\delta=0.3$, then its period is about $T=4.6$s. This is a wind wave.
  As for the swell,
   only if the wave steepness is as low as $\delta=0.03$, its period
  can attain $15$s (Ardhuin et al, 2009).
  Certainly, notice that $A/\delta=\pi/{2k}$ and $T=2\pi/\omega$, this formula
  can be transformed to the form of dissipation relation:
 \begin{eqnarray}
 \omega^2=\frac{\pi^2}{12}gk.
  \end{eqnarray}

With the help of eqn.(2.9), we repeat the above deduction process on the other two stages and get
a combined displacement of piecewise cubic-polynomial form:
\begin{eqnarray}
z(t)=\left\{
\begin{array}{lll}
   \pi \delta g t (T/16-t^2/{3T}),  \quad t\in [0,T/4]\\[2mm]
  \pi \delta g (T/2-t) [T/16-(T/2-t)^2/{3T}],  \enskip t\in [T/4, 3T/4],\\[2mm]
    \pi \delta g (t-T) [T/16-(t-T)^2/{3T}],    \quad t\in [3T/4, T].
\end{array}
\right.
\end{eqnarray}

\subsubsection{Horizontal Motion}

For the horizontal motion, we follow the stage division for the vertical one.
Notice that the trough is on the right side, there should be ${\partial \xi}/{\partial x}<0$ and
$a_x=\delta g$ for the first stage. Corresponding, the horizontal velocity and the displacement
are $u(t)=0+\delta g t$ and $x(t)=0+\delta g t^2/2$.
So the maximum amplitude is
\begin{equation}
L=\frac{1}{2}\delta g \left(\frac{T}{4}\right)^2=\frac{3}{\pi}\cdot \frac{\pi}{96}\delta g T^2=\frac{3}{\pi}A.
\end{equation}
The other three stages can be considered in the same way.
The final displacement relative to the equilibrium at $x_0=0$ is in the form of piecewise quadratic-polynomial:
\begin{eqnarray}
x(t)=\left\{
\begin{array}{lll}
 -3A/\pi+\delta gt^2/2,    \quad t\in [0,T/4]\\[2mm]
 3A/\pi-\delta g (t-T/2)^2/2,  \quad t\in [T/4, 3T/4],\\[2mm]
  -3A/\pi +\delta g(t-T)^2/2,  \quad t\in [3T/4, T].
\end{array}
\right.
\end{eqnarray}

\subsubsection{Motion Synthesis}

 The trajectory of the surface-particle
can be seen as the location shifting of $(x(t),z(t))$ designated by eqn.(2.17) and (2.15).
It follows from \emph{Figure 1} that
the model of piecewise polynomial forms for the horizontal and vertical motions
are very close to the trigonometric forms adopted by the linear wave.
This, at least, indicates that the derived model is acceptable.
It follows from \emph{Figure 2} that
its trajectory is very close to the ellipse $x^2/(3/\pi)^2+z^2=A^2$
which also approximates the circle $x^2+z^2=A^2$ given by the classical linear wave theory.

\section{New Form of Stable Traveling Wave}

\setcounter{equation}{0}
Since the wave fluctuation can be seen as a
coherent movement of a series of surface particles, it is convenient to
express the surface wave (accords with $z_0=0$) in the Lagrangian form $(x(x_0,t), z(x_0,t))$.
Here $x_0$ stands for the ordinary horizontal position and $x(x_0,t)$, $z(x_0,t)$
are the horizontal and vertical displacements of the surface particle relative to the
equilibrium $(x_0,0)$. At first, we describe the instantaneous state of
the wave surface at $t=0$ in the following way.

To incorporate the horizontal location $x_0$ into eqn.(2.17) by a substitution $t=kx_0/\omega$
together with a translational term, it leads to:
\begin{equation}
 x(x_0,0)=\left\{
 \begin{array}{ll}
 x_0-\alpha+\beta (kx_0)^2,\quad kx_0\in[0, \pi/2],\\[2mm]
 x_0+\alpha-\beta\left(kx_0-\pi\right)^2,\enskip kx_0\in[\pi/2,3\pi/2],\\[2mm]
x_0-\alpha+\beta (kx_0-2\pi)^2,\quad kx_0\in[3\pi/2,2\pi],
 \end{array}
   \right.
 \end{equation}
with $\alpha=3A/\pi$, $\beta=\delta g/{2\omega^2}$. For the vertical one,
since our modeling is started from the moment
that the surface particle is at the zero-level with an rising trend,
the wave surface should begin with a trough rather than a crest. So
it follows from eqn.(2.15) that
\begin{equation}
 z(x_0,0)=\left\{
 \begin{array}{ll}
 -\beta kx_0[\gamma-(kx_0)^2]/3,\quad kx_0\in[0, \pi/2],\\[2mm]
 \beta(kx_0-\pi)[\gamma-(kx_0-\pi)^2]/3,\;\;kx_0\in[\pi/2,3\pi/2],\\[2mm]
-\beta(kx_0-2\pi)[\gamma-(kx_0-2\pi)^2]/3,\;kx_0\in[3\pi/2,2\pi]
 \end{array}
   \right.
 \end{equation}
 with $\gamma=3\pi^2/4$.
 Now that the initial wave surface is known, the generation of stable traveling wave form
only requires a substitution of $kx_0$ with $\theta=kx_0-\omega t-2n\pi$ ($n=0,1,2,\cdots$).
The final traveling wave of Lagrangian form is $(x(x_0, t), z(x_0, t))$ designated by (\textsf{Piecewise Polynomial Form}):
\begin{eqnarray}
 &&x(x_0,t)=\left\{
 \begin{array}{ll}
 x_0-\alpha+\beta \theta^2,\quad \theta\in[0, \pi/2],\\[2mm]
 x_0+\alpha-\beta\left(\theta-\pi\right)^2,\enskip \theta\in[\pi/2,3\pi/2],\\[2mm]
x_0-\alpha+\beta (\theta-2\pi)^2,\quad \theta\in[3\pi/2,2\pi],
 \end{array}
   \right.\\
 &&z(x_0,t)=\left\{
 \begin{array}{ll}
 -\beta \theta(\gamma-\theta^2)/3,\quad\theta\in[0, \pi/2],\\[2mm]
 \beta(\theta-\pi)[\gamma-(\theta-\pi)^2]/3,\;\;\theta\in[\pi/2,3\pi/2],\\[2mm]
-\beta(\theta-2\pi)[\gamma-(\theta-2\pi)^2]/3,\;\theta\in[3\pi/2,2\pi]
 \end{array}
   \right.
 \end{eqnarray}
with $\alpha=3A/\pi$, $\beta=\delta g/{2\omega^2}$ and $\gamma=3\pi^2/4$.
These piecewise polynomial forms differ much from the classical trigonometric ones.

\section{New Relations for Wave parameters}
\setcounter{equation}{0} The modeling process yields some brand-new relations for the wave parameters.
First of all, it follows from eqns.(2.6) and (2.13) that
 \begin{eqnarray}
  \omega=\pi\sqrt{\frac{\pi g\delta}{24A}},\qquad k=\frac{\pi\delta}{2A},
  \end{eqnarray}
 here $\delta$ (stands for the mean wave steepness) is incorporated in as an adjusting parameter to convey the diversity of waves.
  With the aid of them, the wave speed can be expressed as:
 \begin{eqnarray}
  c=\frac{\omega}{k}=\sqrt{\frac{\pi g A}{6\delta}}.
  \end{eqnarray}
 This leads to a new understanding on the wave mechanism: \emph{A wave with bigger amplitude or smaller steepness
 travels faster.}

Certainly, the two relations in eqn.(4.1) can be combined to:
\begin{eqnarray}
 \omega^2=\frac{\pi^2}{12}gk\approx 0.82gk,
  \end{eqnarray}
 which is close to the dissipation relation $\omega^2=gk$ adopted by the linear wave.
  For the same wave-number, the frequency of the new model is about $90\%$ that of the linear one.

\section{Two Other Wave Forms}

\setcounter{equation}{0} In view of the matching of
the piecewise polynomial form and the trigonometric form in the description of the particle's motion (\emph{Figure 1}),
we can approximate eqns.(3.3) and (3.4) by
\begin{eqnarray}
\left\{
\begin{array}{ll}
x(x_0, t)=x_0-(3/\pi) A\cos(kx_0-\omega t),\\[2mm]
z(x_0, t)=-A\sin(kx_0-\omega t).
\end{array}
\right.
\end{eqnarray}
Relative to the piecewise polynomial form, this version of surface wave is more convenient to use.
To incorporate the factor $e^{kz_0}$ together with a translation on the equilibrium
from $(x_0,0)$ to $(x_0,z_0)$, it results in (\textsf{Trigonometric Form}):
\begin{eqnarray}
\left\{
\begin{array}{ll}
x(x_0,z_0, t)=x_0-(3/\pi) Ae^{kz_0}\cos(kx_0-\omega t),\\[2mm]
z(x_0,z_0, t)=z_0-Ae^{kz_0}\sin(kx_0-\omega t),
\end{array}
\right.
\end{eqnarray}
which can be used to describe the wave motion in deep-water. To compare with eqn.(1.10) we see
this form is very close to that of the Gestner wave.

To meet the using habit of wave study we also provide a form of $z=\xi(x,t)$ (\textsf{Free Surface Form}) below:
  \begin{eqnarray}
  z=-A\sin\left[\theta+1.5\delta\cos\left(\theta+1.5\delta\cos{\theta}\right)\right]
\end{eqnarray}
 with $\theta=kx-\omega t$, which can be seen as a further approximation to the trigonometric form.
 Particularly, in case $\delta\to 0$ it degenerates to the linear wave
 $z=-A\sin(kx-\omega t)$.
 Here the substitution $kA=\pi\delta/2$ and the second-order approximation below are used:
\begin{eqnarray}
x_0&=&x+(3/\pi) A\cos(kx_0-\omega t)\nonumber\\
&=&x+(3/\pi) A\cos\left\{k\left[x+(3/\pi) A\cos(kx_0-\omega t)\right]-\omega t\right\}\nonumber\\
&\approx&x+(3/\pi) A\cos\left[ kx-\omega t+(3/\pi)kA\cos(kx-\omega t)\right].\nonumber
\end{eqnarray}

To show the matching of the free surface form with the other two forms, a numerical test is done in \emph{Figure 3}.
It shows that the second-order approximation is well for the lower slope case. But for the too high
case ($\delta=0.6$ accords with wave-breaking)
 it deviates much from the original piecewise polynomial form.
Hence, this form is only for non-breaking case and it is not valid for wave-breaking arguments.
Certainly, for the non-breaking case if it is needful this kind of approximation can be extended to a higher order.
Each time the phase angle is corrected by a periodic one with magnitude $1.5\delta$.

In addition, due to the horizontal motion of the surface particle,
the asymptotic expansion technique for Stokes wave is awkward in approximating the free surface $z=\xi(x,t)$.
But in fact, the second-order approximation of our nested form is enough to reflect
the crest-trough asymmetry.

\section{New Wave-Breaking Criteria}
\setcounter{equation}{0}
The study of the physical and dynamical characteristics of gravity waves on the sea surface
while they break and the subsequent foam activity and formation of drop-spray clouds are amongst the
major problems facing modern satellite oceanology, physics of the ocean-atmosphere interaction,
and oceanic engineering (Sharkov, 2007).

\subsection{Review on Wave-Breaking Problem}

The Breaking phenomenon is usually associated with steep waves at the sea surface.
To divide from the research objects, there are two classes of criteria for wave-breaking:
 the first is related to the characteristic of the
surface elevation and the second is related to observations of air entrainment,
whitecaps or ambient noise. Here only the first class is concerned.

More than one hundred years ago, Stokes had brought forth a finite-amplitude wave theory
and established a set of criteria for wave-breaking (Massel, 2007):\\
\emph{(a) The particle's velocity at the wave crest $u_c$ equals to the
wave speed $c$ [\textsf{kinematic criterion}];\\
(b) The wave crest attains a sharp point with an angle
of $120^{\circ}$, that is, the upper wave part accords with a mean slant angle $\alpha=30^{\circ}$ [\textsf{angle criterion}];\\
(c) The wave steepness in terms of $\varepsilon^*=H/\lambda$ approximates $1/7$,
here $H=2A$ stand for the wave height [\textsf{steepness criterion}];\\
(d) The particle's acceleration at the crest estimated under the polar-coordinate frame is
 $a_c=g/2$, here $g$ is the gravity acceleration [\textsf{acceleration criterion}].}

These four criteria are seen as classical ones nowadays. Among all of them, the kinematic criterion is the fundamental one which accords with
the fact: \emph{this position represents
the stagnation point of the fluid-particle's streamline relative to the wave form}.
Downstream from this point, fluid particles tend to escape from the water
surface. They can either be ejected into air to become droplets or curl down to
trap air into water. The other three criteria are merely limiting approximations to
the steepest wave with a preset velocity potential of polar-coordinate form $\phi(r,\theta)=Br^n\sin(n\theta)$
in the crest region, where $B$ and $n$ are the coefficients to be evaluated.
Certainly, the different choices of velocity potential
with different simplicities of the equations imply the different criteria
and the successors have given a variety of modifications for them.
For example, in 1977 Longuet-Higgins modified the slant angle and the acceleration criterions
to be $\alpha=30.37^{\circ}$ and $a_c=0.388g$ (Massel, 2007).

Essentially, the Stokes wave can not be directly used for wave-breaking investigation
due the form $z=\xi(x,t)$. So the indirect approach is adopted. However, the wave
of Lagrangian form $(x(x_0,t),z(x_0,t))$ can be directly used due to the advantage
of reflecting the particle's horizontal motion.
In the following we abandon the velocity-potential approach
and restudy this problem with the newly derived model.
To make comparison between the piecewise polynomial form and the trigonometric form,
we see the first one with lower smoothness is more preferable,
 after all, the wave with a too sharp crest is not smooth.

\subsection{Modeling the New Breaking Criteria}

Relative to the linear wave $\xi(x,t)=-A\sin(kx-\omega t)$ and the Stokes wave in eqns.(1.6) (its equilibrium is adjusted to $z=0$ with
adding $-\varepsilon A/2$), the new model has more distinct physical meaning and it accords better with
the actual one which has a sharp crest and a flat trough
(see \emph{Figure 4}).
In fact, this asymmetry roots in the horizontal displacement of the surface particle. According to eqn.(3.3),
in one wavelength $\lambda$ the crest part and the trough part possess lengths
$$\frac{\lambda}{2}-2\cdot\frac{3}{\pi}A=2(\frac{1}{\delta}-\frac{3}{\pi})A, \quad
\frac{\lambda}{2}+2\cdot\frac{3}{\pi}A=2(\frac{1}{\delta}+\frac{3}{\pi})A$$
separately.
So, a bigger $\delta$ implies a sharper crest and this characteristic
can be used to forecast the wave-breaking.

Notice that the kinematic criterion given by Stokes is a generally accepted one,
we take it as a theoretical basis and give further deduction.
We note here that the kinematic criterion can not be directly used since
it involves in determining the wave speed and surface-particle's velocity at the wave crest.

\subsubsection{For the Original Wave of Piecewise Polynomial Form}

On the one hand, it follows from the modeling process that the surface particle at the wave crest possesses a
zero vertical velocity and a maximum horizontal velocity:
\begin{equation}
u_c=\delta g \frac{T}{4}=\delta g\sqrt{\frac{6A}{\pi g\delta}}=\sqrt{\frac{6Ag\delta}{\pi}},
\end{equation}
here the formula (2.13) for $T$ is used. On the other hand, it follows from eqn.(4.2) that
$c=\sqrt{\pi gA/{6\delta}}$. To meet the kinematic criterion it requires
$u_c\geq c$ which leads to
\begin{equation}
\delta\geq \frac{\pi}{6}.
\end{equation}
This is a steepness type of criterion which is higher than the one given by Stokes
since, at this time, the critical case reads $\varepsilon^*=H/\lambda=\delta/2=\pi/12>1/7$.

The above steepness type of criterion can be also transformed to the angle type.
Notice that the wave steepness $\delta=4A/\lambda$ is actually the averaged wave slope,
these two type are equivalent to each other. Hence,
 the steepest wave should possess a mean slant angle:
 \begin{equation}
 \alpha=\arctan(\pi/6)\approx 27.6^{\circ}
 \end{equation}
which is slightly lower than $30^{\circ}$ and $30.37^{\circ}$ given by Stokes and Longuet-Higgins, respectively.

In case the asymmetry property is considered,
the crest part should possess a critical wave slope
 \begin{equation}
\delta^*=\frac{A}{\lambda/4-3A/\pi}=\frac{\pi\delta}{\pi-3\delta}
=\frac{\pi\cdot{\pi}/{6}}{\pi-3\cdot{\pi}/{6}}=\frac{\pi}{3}.
 \end{equation}
which results in $\alpha^*=\arctan{\delta^*}\approx 46.3^{\circ}$. \emph{This indicates that
a wave with front angle bigger than $46.3^{\circ}$ must break.}
It has more guiding significance than the mean one, after all, more often than not
the superposition of two non-breaking wave may result in a breaking crest (see \emph{Figure 7})
which can not be well scaled by other approaches.
As for the trough part, the critical slope should be
$\delta_*=A/(\lambda/4+3A/\pi)=\pi\delta/(\pi+3\delta)=\pi/9$ which accords with
a slant angle $\alpha_*=\arctan{\delta_*}\approx 19.2^{\circ}$.

We note that the acceleration criterion is not preferred for our model.
The reason is that, the horizontal acceleration of the surface particle has no definition at the crest.
 Before the particle attains the crest it is $a_x=\delta g$
 and after the crest it turns to $a_x=-\delta g$.
 But their magnitudes $|a_x|=\delta g=\pi g/6\approx 0.52 g$ approximate $a_c=0.5g$ given by Stokes.
 Besides the horizontal acceleration, there is also a
vertical one with $a_z=-\pi^2 g/12\approx -0.82 g$ at the crest.

\subsubsection{For the Approximated Wave of Trigonometric Form}

If the wave of trigonometric form is adopted,
the higher-order smoothness of it may postpone the occurrence of wave-breaking.
It follows from eqn.(5.1) that
\begin{eqnarray}
u=\frac{dx}{dt}=-\frac{3}{\pi}A\omega\sin(kx_0-\omega t)\nonumber
\end{eqnarray}
which yields a maximum horizontal velocity at the crest with $x_0=0$ and $t=T/4$:
\begin{eqnarray}
u_c=\frac{3}{\pi}A\omega =\frac{3}{\pi}A\pi\sqrt{\frac{\pi g\delta}{24A}}=
\sqrt{\frac{3\pi gA\delta}{8}}.
\end{eqnarray}
To ensure $u_c\geq c$ it only requires
\begin{equation}
\delta\geq \frac{2}{3}
\end{equation}
which accords with a critical mean-slope angle $\alpha=33.7^{\circ}$.
For this case, the critical front slope and front angle are $\delta^*=2\pi/3(\pi-2)$
and $\alpha^*=61.4^{\circ}$.
Relative to $46.3^{\circ}$ derived from the original form,
to approximate the wave with trigonometric function
may result in an error $15.1^{\circ}$ to the critical front angle.
Certainly, the original piecewise polynomial form is also an approximation to the actual wave.
These results are left to be checked by observations.

\subsection{Numerical Tests on Wave-Breaking}

It follows from the previous section that $\delta=\pi/6$ is the theoretical critical mean-slope.
So in case $\delta>\pi/6$ the wave should break and on the contrary
it doesn't. In the following we check it by numerical approach.

Firstly, we make a simulation on the effect of wave slope.
It follows from \emph{Figure 5} that a bigger wave slope accords with a
sharper crest. From the sub-figure \textsf{d} we see the breaking characteristic is very
obvious in case $\delta=0.8$ (indicated by the small curl at the crest).
In fact, the breaking has already occurred just bigger than $\pi/6$.
To find the exact critical value of $\delta$ it needs zooming in the crest part.
From \emph{Figure 6} we see the breaking does not occur for the
case $\delta\leq \pi/6$ ($\approx 0.524$) and does surely occur at $0.525$.
This test indicates that $\delta=\pi/6$ is indeed a critical mean-slope.
In another word, the critical front slope should be $\delta^*=\pi/3$
and it can be taken as a feasible wave-breaking criterion.
The numerical results in \emph{Figure 5} also imply the smaller the wave slope
 the closer the curve to the sinusoidal one. This reflects the
consistency of the new model and the linear model.

In addition, the breaking criteria $\delta=2/3$ for the trigonometric form can be also tested in the same way.
 \emph{Figure 7} shows the superposition of two waves. It is easy to see from \textsf{c} that
 the superimposed wave has many breaking crests, though these are the non-breaking ones with $\delta=0.3$ and $0.5$
 separately.

\section{Improvement on the New Wave Model}
\setcounter{equation}{0}
Beside the advantages mentioned in \emph{Section 6.2}, there is still a shortcoming for this new model: its particle trajectory is a closed one.
To keep up with the Stokes wave, it should be able to reflect the
wave drift.
In the following, we improve the model by substituting
 the mean-slope $\delta$ with two local slopes
  for the crest and trough parts defined in \emph{Section{6.2.1}}:
 \begin{eqnarray}
 \delta^*=\frac{\pi\delta}{\pi-3\delta},\quad \delta_*=\frac{\pi\delta}{\pi+3\delta}.
 \end{eqnarray}
Notice that this change mainly affects the horizontal motion
and there is little influence on the vertical one,
we pay our attention to the horizontal remodeling.

For this case, the horizontal acceleration in eqn.(2.7) is substituted by
\begin{eqnarray}
a_x=\left\{
\begin{array}{ll}
\delta^* g e^{kz_0},\quad z>0,\; {\partial\xi}/{\partial x}<0,\\[2mm]
-\delta^* g e^{kz_0}, \quad z>0,\; {\partial\xi}/{\partial x}>0,\\[2mm]
-\delta_* g e^{kz_0}, \quad z<0,\; {\partial\xi}/{\partial x}>0,\\[2mm]
\delta_* g e^{kz_0}, \quad z<0,\; {\partial\xi}/{\partial x}<0.
\end{array}
\right.
\end{eqnarray}
To repeat the deduction process in \emph{Section 2.3} it yields
a reformed traveling wave (\textsf{Improved Piecewise polynomial Form}):
\begin{eqnarray}
 &&x(x_0,t)=\sigma_1 n+\left\{
 \begin{array}{ll}
x_0-\alpha_1+\beta_1 \theta^2,\quad \theta\in[0, \pi/2],\\[2mm]
x_0+\alpha_1-\beta_1\left(\theta-\pi\right)^2,\enskip \theta\in[\pi/2,\pi],\\[2mm]
x_0+\alpha_1-\beta_2\left(\theta-\pi\right)^2,\enskip \theta\in[\pi,3\pi/2],\\[2mm]
x_0-\sigma_2+\beta_2 (\theta-2\pi)^2,\;\; \theta\in[3\pi/2,2\pi],
 \end{array}
   \right.\\
 &&z(x_0,t)=\left\{
 \begin{array}{ll}
 -\beta_0 \theta(\gamma-\theta^2)/3,\quad\theta\in[0, \pi/2],\\[2mm]
 \beta_0(\theta-\pi)[\gamma-(\theta-\pi)^2]/3,\;\;\theta\in[\pi/2,3\pi/2],\\[2mm]
-\beta_0(\theta-2\pi)[\gamma-(\theta-2\pi)^2]/3,\;\theta\in[3\pi/2,2\pi]
 \end{array}
   \right.
 \end{eqnarray}
with $\alpha_1=3A\delta^*/{\pi\delta}$, $\alpha_2=3A\delta_*/{\pi\delta}$,
$\beta_0=\delta g/{2\omega^2}$, $\beta_1=\delta^* g/{2\omega^2}$, $\beta_2=\delta_* g/{2\omega^2}$,
$\gamma=3\pi^2/4$, $\sigma_1=2(\alpha_1-\alpha_2)$, $\sigma_2=2\alpha_2-\alpha_1$
and $\theta=kx_0-\omega t-2n\pi$ ($n=0,1,\cdots$).

With these revisions, the approximations in eqn.(5.3) become
(\textsf{Improved Trigonometric Form}):
\begin{eqnarray}
&&x(x_0,z_0, t)=x_0+e^{kz_0}\cdot\left\{
 \begin{array}{ll}
\sigma_1 n-\alpha_1 \cos{\theta},\;\; \theta\in[0,\pi],\\[2mm]
\sigma_1 (n+1/2)-\alpha_2 \cos{\theta},\; \theta\in[\pi,2\pi],
 \end{array}
 \right.\\
&&z(x_0,z_0, t)=z_0-Ae^{kz_0}\sin{\theta}.
\end{eqnarray}
For convenience of using, further approximation can be done on the term $\sigma_1 n\approx U_d t$, here $U_d$ is the
drift velocity defined in the next section.

\section{Wave Drift Induced by the New Model}
\setcounter{equation}{0}
For simplicity, the improved trigonometric forms in eqn.(7.5) and (7.6) are used here.
It follows from \emph{Figure 8} that the particle's trajectory
is no longer a closed one and there is a distinct drift.

In fact, for a particle with equilibrium at depth $z_0$, on each period $T$
its drift distance is $\sigma_1 e^{kz_0}$. So the drift velocity is:
 \begin{eqnarray}
 U_d=\frac{\sigma_1 e^{kz_0}}{T}=\frac{6A(\delta^*-\delta_*)}{\pi\delta}\sqrt{\frac{\pi g\delta}{96A}} e^{kz_0}
 =\frac{3\delta \sqrt{6\pi gA\delta}}{2(\pi^2-9\delta^2)} e^{kz_0},
 \end{eqnarray}
 which only relies on $A, \delta$ and $z_0$.
 Relative to the Stokes drift $U_s=\varepsilon^2 c_s e^{2kz_0}$ ($c_s$ stands for the wave speed in its own frame),
 the newly derived one yields a much lower wave drift ($\leq 1.6$m/s for $\delta\leq 0.5$)
 whose magnitude is more rational (see \emph{Figure 9}).
 The surface drift estimated by Stokes ($=5.5$m/s for $\delta=0.5$) is too strong to meet
 the common sense. It is rare to observe a current of several-knots at sea, not to say the drift
 for the particle's trajectory.

\section{Summaries and Discussions}
\setcounter{equation}{0}

By modeling approach we have deduced a new model for the ocean surface wave which
differs from the classical ones given by Airy (the linear wave model), Stokes (the nonlinear wave model) and Gerstner (the cycloid wave model).
The corresponding dissipation relation is
\begin{equation}
\omega^2=\frac{\pi^2}{12}gk.
\end{equation}
In terms of amplitude $A$ and slope $\delta$, the wave parameters can be rewritten as
 \begin{eqnarray}
  \omega=\pi\sqrt{\frac{\pi g\delta}{24A}},\quad k=\frac{\pi\delta}{2A}, \quad c=\sqrt{\frac{\pi g A}{6\delta}}.
  \end{eqnarray}
 These kind of relations have more guiding significants for the field observations.
 The new understandings on wave mechanism are as follows:\\
  \emph{(1) A wave with bigger amplitude or smaller slope should travel faster
 (The common sense is that: lower frequency or
 lower wave-number implies faster);\\
 (2) The critical mean-slope and front-slope for wave-breaking are $\delta=\pi/6$ and $\delta^*=\pi/3$
 which accord with slant angles $27.6^{\circ}$ and $46.3^{\circ}$ separately (Stokes's front angle is $30^{\circ}$);\\
  (3) The new formula indicates that the magnitude of wave drift should be smaller than that of the known Stokes drift.}

Besides the wave-breaking problem, the formulas in eqn.(9.2), especially the first one, can be also used
to study other problems, their guiding significant is self-evident.

In case the wave is under a developing process, the first relation can give a well guidance
to its variation. That is to say, along with the increasing of amplitude $A$,
its period increases in a clear way with $T=\sqrt{96A/{\pi g\delta}}$.
It follows from \emph{Figure 10} that if the wave amplitude is increased from $4$m to $5$m
then the corresponding period should increase from $8.0$s to $8.8$s provided that the slope maintains unchanged with $\delta=0.2$.

The first formula can be also used in a reverse way. Notice that the amplitude
is $A=\pi^3 g\delta/{24\omega^2}$, we can express the wave energy (the factor $\rho g$ is dropped) by
 \begin{eqnarray}
 E=\frac{1}{2} A^2=\frac{\pi^6  g^2}{1152}\cdot\frac{\delta^2}{\omega^4}.\nonumber
  \end{eqnarray}
Therefore, the energy relies not only on the frequency but also on the steepness, that is, $E\propto \delta^2 \omega^{-4}$.
This theoretical result is helpful for improving the existing wave-spectrum theories
whose main usages are to find the relations between $\omega$ and $A$ by considering the stochastic ocean surface
as a superposition of a series of linear waves.
This formula is close to the theoretical Zakharov-Filonenko spectrum with $E\propto \omega^{-4}$ (Cavaleri et al, 2007);
the empirical Pierson-Moscowitz and JONSWAP spectrums with $E\propto \omega^{-5}$ (Janssen, 2009)
and the empirical Neumann spectrum with $E\propto \omega^{-6}$ (Wen and Yu, 1984).
Certainly, to reconstruct the wave-spectrum theory needs investigating the distribution
of wave-slope $\delta$ at sea. It requires further research.

There is a necessity to compare the new model with the classical ones.
As the modeling approach concerned, the traditional ones given by Airy, Stokes and Gerstner
take the conjectured wave forms as the preconditions or
approximate it with asymptotic expansion technique.
What is more, the unknown water pressures are given as corollaries in the last.
Our approach is an inverse one: the conjecture is done on the pressure, to deduce the wave form is taken
as an object. The new idea is to consider the local Lagrangian motion of the surface particle
with Newton's second law.

The new model is originally developed in a piecewise polynomial form
which has distinct physical meaning. It accords well with the actual one which has a sharp crest and a flat trough.
Its approximated trigonometric form is very close to the Gerstner wave; Its approximated free-surface form
takes the linear wave as a limiting case; The nested fashion for the free-surface form is superior to
the asymptotic expansion technique adopted by the Stokes wave. In addition, the improved model can also reflect
the wave drift and its magnitude is more rational than the one given by Stokes.

This new model together with its brand-new relations
are left to be checked by observations.

\vskip 4mm
\noindent\textbf{Acknowledgments.} We thank the supports from
the Fund of Oceanic Telemetry Engineering and Technology Research Center,
State Oceanic Administration of China (No.2013005).


\newpage

\begin{figure}
  \centerline{\includegraphics[height=2.6in,width=3.5in]{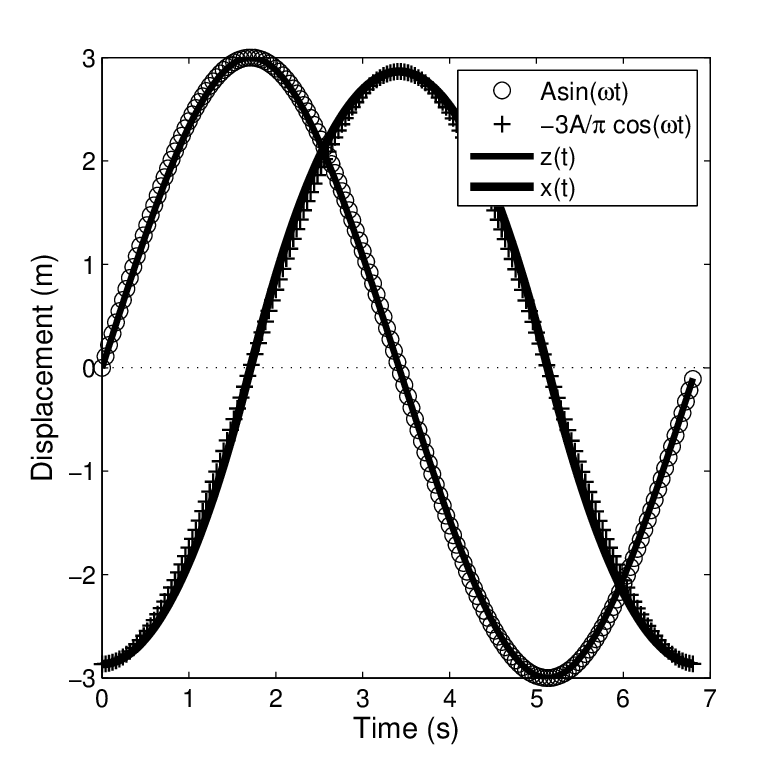}}
  \caption{\footnotesize The time-variations of the horizontal and vertical displacements for the surface particle
with $A=3$, $\delta=0.2$.
For comparison, the trigonometric functions $A\sin(\omega t)$ and $-3A/\pi\cos(\omega t)$
with $\omega=2\pi \sqrt{\pi g\delta/{96A}}$ are also drawn.}
\end{figure}

\begin{figure}
\centerline{\includegraphics[height=2.6in,width=3.5in]{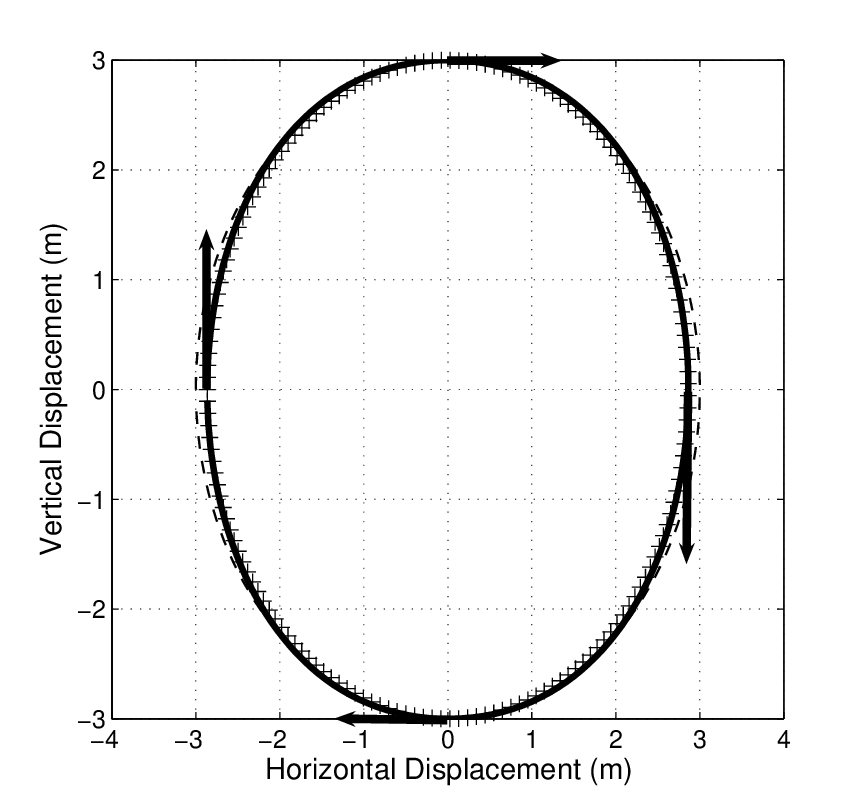}}
\caption{\footnotesize The trajectory of the surface particle with $A=3$, $\delta=0.2$ (---).
 For comparison the ellipse $x^2/(3/\pi)^2+z^2=A^2$ (++) and the circle
 $x^2+z^2=A^2$ (- -) are also drawn.
 Here the four arrows depict the velocities
at the key points.}
\end{figure}
\begin{figure}
\centerline{\includegraphics[height=2.5in,width=4in]{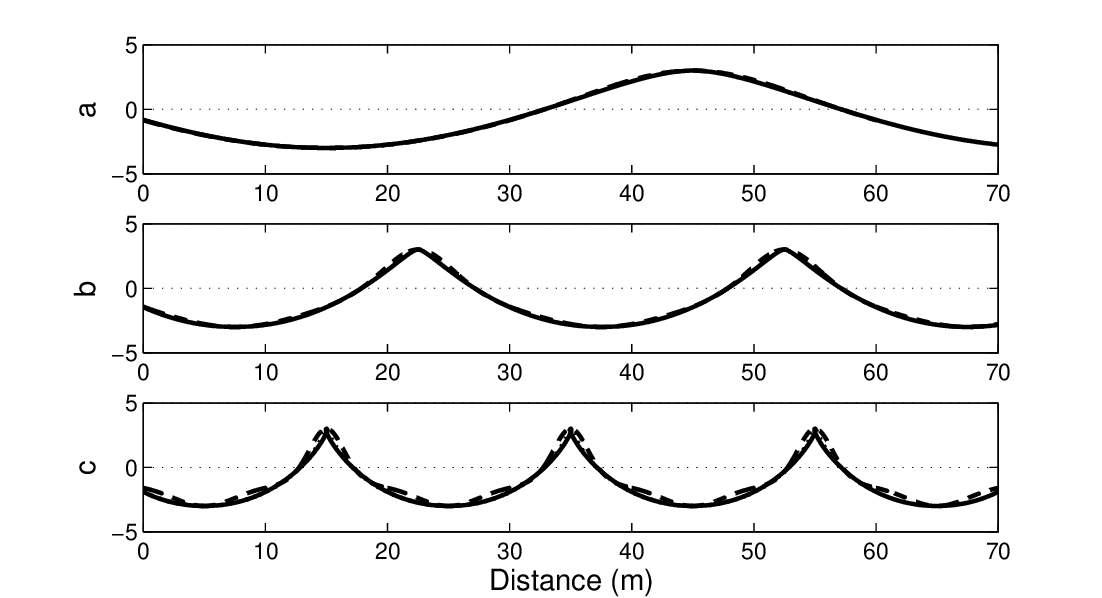}}
\caption{\footnotesize Comparison between the free surface form (- -),
the piecewise polynomial form (---) and the trigonometric form ($\cdots$).
From \textsf{a} to \textsf{c} the values of $\delta$ are $0.2, 0.4$ and $0.6$ separately.}
\end{figure}
\begin{figure}
\centerline{\includegraphics[height=2in,width=4.4in]{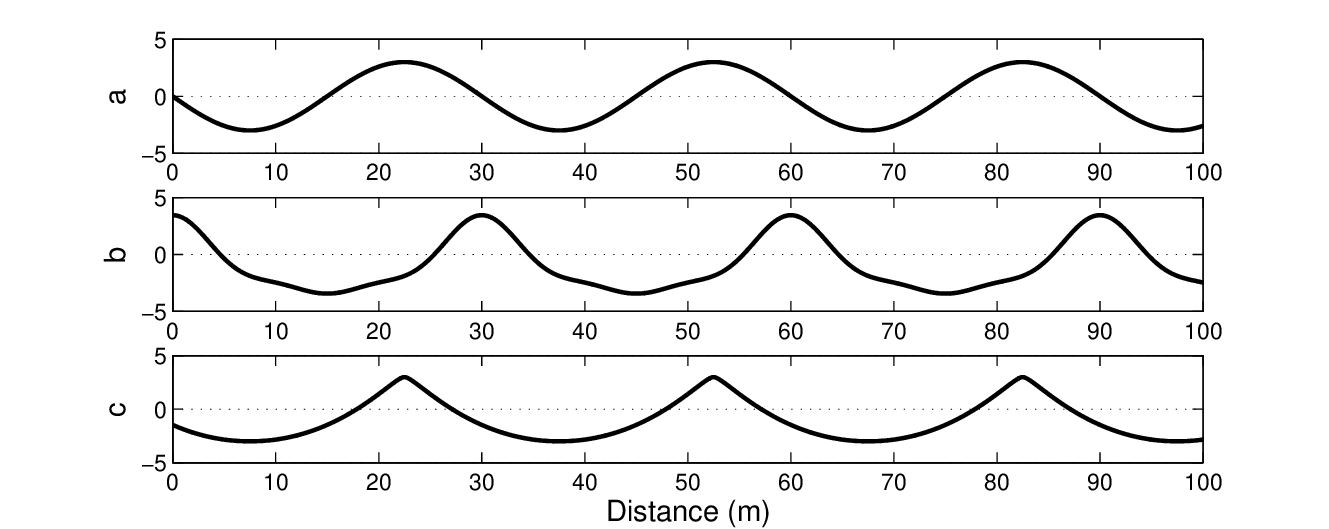}}
\caption{\footnotesize Comparison between three wave models with $A=3$m, $\delta=0.4$
and $k=\pi \delta/{2A}$ for the case $t=0$. a: the linear wave $-A\sin(kx)$;
b: the third-order Stokes-wave; c: the new wave of piecewise polynomial form.}
\end{figure}
\begin{figure}
\centerline{\includegraphics[height=2.2in,width=4in]{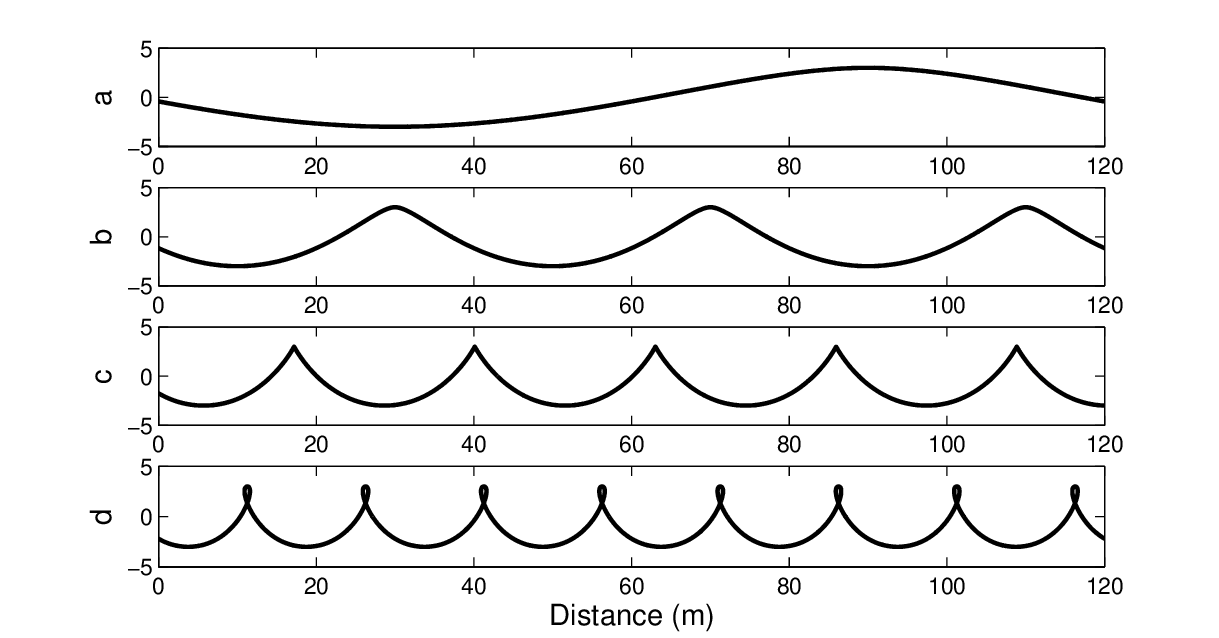}}
\caption{\footnotesize The variation of the wave surface in piecewise polynomial form
 along the mean wave slope. From \textsf{a} to \textsf{d}
 the values of $\delta$ are $0.1, 0.3, \pi/6$ and $0.8$ separately.}
\end{figure}
\begin{figure}
\centerline{\includegraphics[height=1.8in, width=4in]{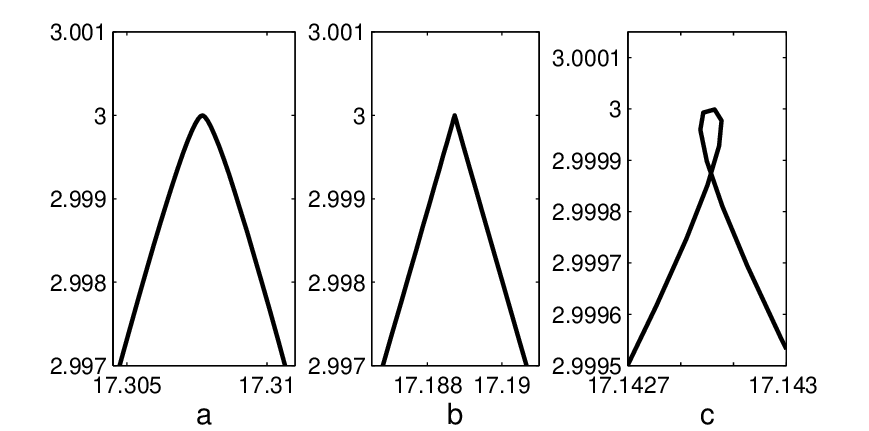}}
\caption{\footnotesize The detailed crest part
for the wave surface in piecewise polynomial form
 along the mean wave slope. From \textsf{a} to \textsf{c}
 the values of $\delta$ are $0.520, \pi/6$ and $0.525$ separately. }
\end{figure}
\begin{figure}
\centerline{\includegraphics[height=2.5in, width=3.5in]{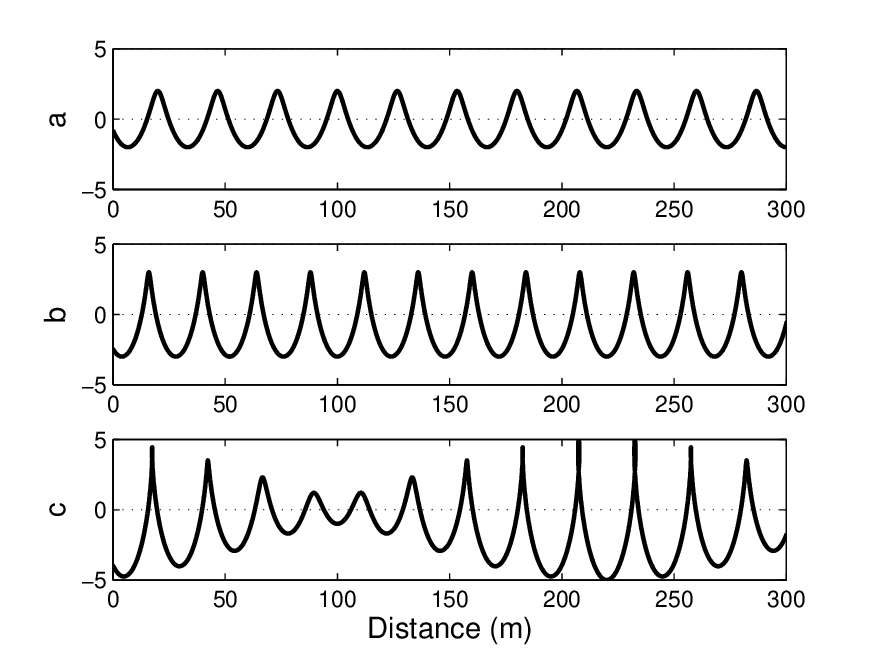}}
\caption{\footnotesize Two non-breaking waves and their superposition. \textsf{a}: the wave of trigonometric form
with $A=2$m and $\delta=0.3$; \textsf{b}: the wave of trigonometric form
with $A=3$m and $\delta=0.5$; \textsf{c}: the superposition of the two waves. }
\end{figure}
\begin{figure}
\centerline{\includegraphics[height=2.0in,width=3.2in]{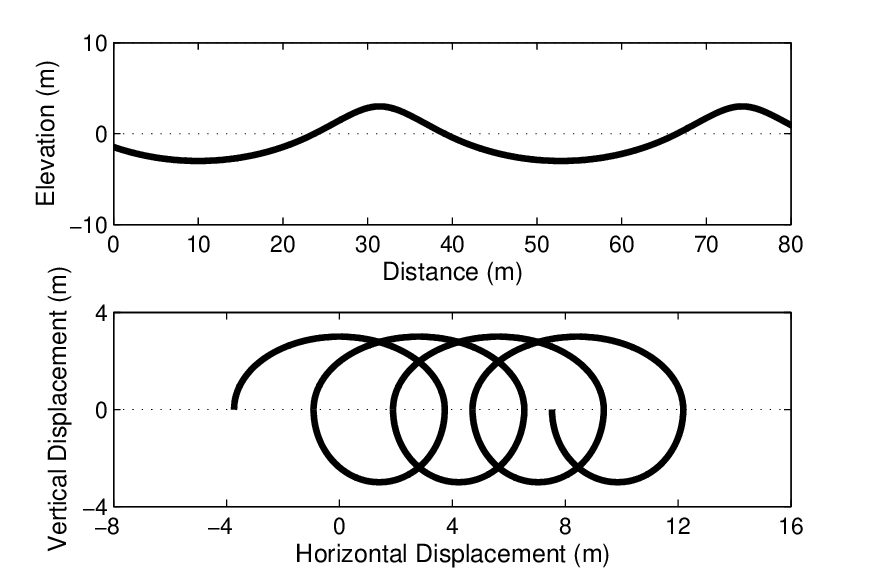}}
\caption{\footnotesize The wave surface (upper sub-figure) and the trajectory for a particle at the surface (lower sub-figure)
respect to the improved trigonometric form with $A=3$m, $\delta_0=0.3$ and $z_0=0$.}
\end{figure}
\begin{figure}
\centerline{\includegraphics[height=2.2in,width=3.0in]{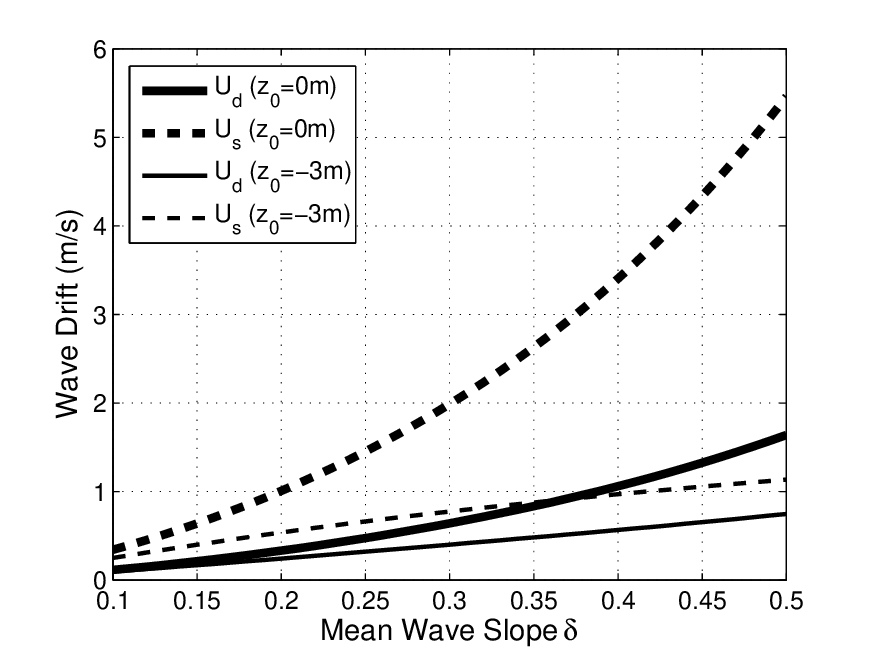}}
\caption{\footnotesize Comparison between the newly derived drift $U_d$ and the Stokes drift
 with equal $A$ ($=3$m) and $k=\pi\delta/{2A}$.}
\end{figure}
\begin{figure}
\centerline{\includegraphics[height=2.2in, width=3in]{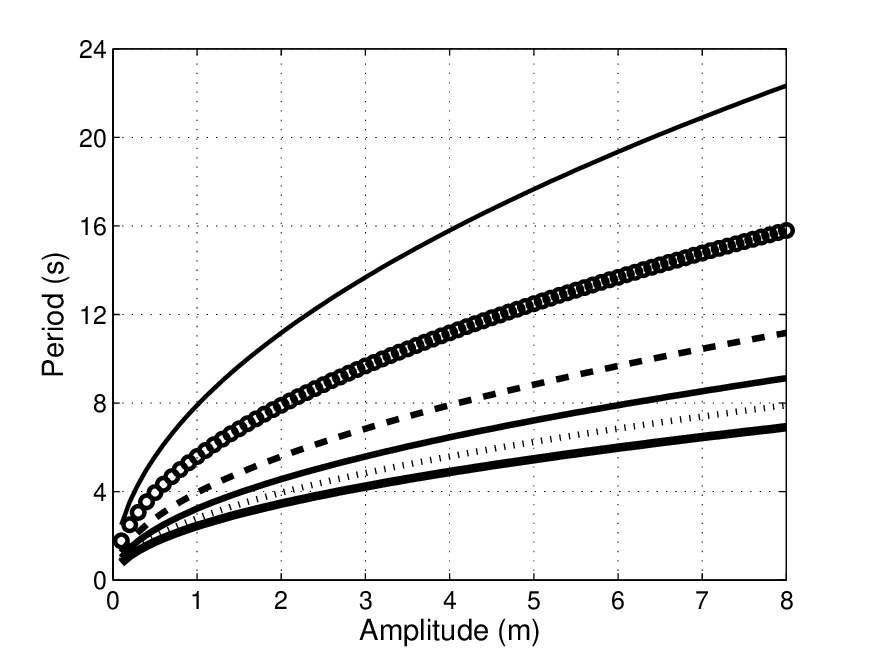}}
\caption{\footnotesize The variation of wave period along with the increasing of amplitude. From
the upper to the lower the slopes are $0.05, 0.1, 0.2, 0.3, 0.4$ and $\pi/6$ separately.  }
\end{figure}

\end{document}